\documentclass[11pt]{article}

\usepackage[utf8]{inputenc}
\usepackage[vmargin=1.25in]{geometry}
\usepackage{amsmath,amssymb}
\usepackage{mathrsfs}
\usepackage[square]{natbib}
\usepackage{graphicx}
\usepackage[colorlinks,citecolor=green,linkcolor=blue,bookmarks=false,hypertexnames=true]{hyperref}

\newtheorem{remark}{Remark}

\renewcommand{\d}[1]{\ensuremath{\operatorname{d}\!{#1}}}
\newcommand{\D}[1]{\ensuremath{\operatorname{D}\!{#1}}}

\newcommand{\del}[2]{\frac{\delta{#1}}{\delta{#2}}}
\def\const{\ensuremath{\operatorname{const}}}

\def\Id{\ensuremath{\operatorname{Id}}}

\def\SDiff{\ensuremath{\SDiff}}

\DeclareMathOperator{\neigh}{neighbor}

\makeatletter
\newcommand*{\bdot}{}% Check if undefined
\DeclareRobustCommand*{\bdot}{%
  \mathbin{\mathpalette\bdot@{}}%
}
\newcommand*{\bdot@scalefactor}{.5}
\newcommand*{\bdot@widthfactor}{1.15}
\newcommand*{\bdot@}[2]{%
  \sbox0{$#1\vcenter{}$}% math axis
  \sbox2{$#1\cdot\m@th$}%
  \hbox to \bdot@widthfactor\wd2{%
    \hfil
    \raise\ht0\hbox{%
      \scalebox{\bdot@scalefactor}{%
        \lower\ht0\hbox{$#1\bullet\m@th$}%
      }%
    }%
    \hfil
  }%
}
\makeatother

\DeclareFontFamily{U}{mathx}{\hyphenchar\font45}
\DeclareFontShape{U}{mathx}{m}{n}{
      <5> <6> <7> <8> <9> <10>
      <10.95> <12> <14.4> <17.28> <20.74> <24.88>
      mathx10
      }{}
\DeclareSymbolFont{mathx}{U}{mathx}{m}{n}
\DeclareMathSymbol{\btimes}{1}{mathx}{"91}

\usepackage{pict2e}
\makeatletter
\DeclareRobustCommand{\intprod}{%
  \mathbin{\mathpalette\int@prod{(0.1,0)(0.9,0)(0.9,0.8)}}%
}
\DeclareRobustCommand{\intprodr}{%
  \mathbin{\mathpalette\int@prod{(0.1,0.8)(0.1,0)(0.9,0)}}}
\newcommand{\int@prod}[2]{%
  \begingroup
  \sbox\z@{$\m@th#1+$}%
  \setlength\unitlength{\wd\z@}%
  \begin{picture}(1,1)
  \roundcap
  \polyline#2
  \end{picture}%
  \endgroup
}
\makeatother

\def\x{\mathbf x}

\def\u{\mathbf u}
\def\v{\mathbf v}
\def\w{\mathbf w}

\def\p{\psi}
\def\ps{\psi}
\def\xx{q}

\def\n{\hat{\mathbf n}}

\def\H{\mathscr H}

\def\C{\mathscr C}
\def\U{\mathscr U}
\def\V{\mathscr V}
\def\W{\mathscr W}

\usepackage{titling}
\usepackage{array}
\preauthor{\begin{center}
    \large \lineskip .75em%
        \begin{tabular}[t]{>{\centering\arraybackslash}p{.45\textwidth}}}
        \postauthor{\end{tabular}\par\end{center}}
\makeatletter
\renewcommand\and{%                  % \begin{tabular}
  \end{tabular}%
  \hfill
  \begin{tabular}[t]{>{\centering\arraybackslash}p{.45\textwidth}}}%   % \end{tabular}
  \makeatother

\begin{document}

\title{Carriers of \emph{Sargassum} and mechanism for coastal
inundation in the Caribbean Sea}

\author{F.\ Andrade-Canto\\ Departamento de Observaci\'on y Estudio
de la Tierra, la Atm\'osfera y el Oc\'eano\\ El Colegio de la
Frontera Sur\\ Chetumal, Quintana Roo, Mexico\\
fernando.andrade@ecosur.mx\and F.J.\ Beron-Vera\\ Department of
Atmospheric Sciences\\ Rosenstiel School of Marine \& Atmospheric
Science\\ University of Miami\\ Miami, Florida, USA\\ fberon@miami.edu
\and G.J.\ Goni\\Atlantic Oceanographic and Meteorological Laboratory\\
National Oceanic \& Atmospheric Administration\\ Miami, Florida,
USA\\ gustavo.goni@noaa.gov \and D.\ Karrasch\\ Technische Universit\"at
M\"unchen\\ Zentrum Mathematik\\ Garching bei M\"unchen, Germany\\
karrasch@ma.tum.de \and M.J.\ Olascoaga\\ Department of Ocean
Sciences\\ Rosenstiel School of Marine \& Atmospheric Science\\
University of Miami\\ Miami, Florida, USA\\ jolascoaga@miami.edu
\and J.\ Tri\~nanes\thanks{Also at Cooperative Institute for Marine
\& Atmospheric Studies, University of Miami, Miami, Florida, USA
and Departamento de Electr\'onica y Computaci\'on, Universidade de
Santiago de Compostela, Santiago, Spain.}\\ Atlantic Oceanographic
and Meteorological Laboratory\\ National Oceanic \& Atmospheric
Administration\\ Miami, Florida, USA\\ joaquin.trinanes@noaa.gov}
\date{Started: September 23, 2021. This version: \today.\vspace{-0.25in}}
\maketitle

\begin{abstract}
  We identify effective carriers of \emph{Sargassum} in the Caribbean
  Sea and describe a mechanism for coastal choking.  Revealed from
  satellite altimetry, the carriers of \emph{Sargassum} are mesoscale
  eddies (vortices of 50-km radius or larger) with coherent material
  (i.e., fluid) boundaries.  These are observer-independent---\emph{unlike}
  eddy boundaries identified with instantaneously closed streamlines
  of the altimetric sea-surface height field---and furthermore
  harbor finite-time attractors for networks of elastically connected
  finite-size buoyant or ``inertial'' particles dragged by ocean
  currents and winds, a mathematical abstraction of \emph{Sargassum}
  rafts.  The mechanism of coastal inundation, identified using a
  minimal model of surface-intensified Caribbean Sea eddies, is
  thermal instability in the presence of bottom topography.
\end{abstract}

%\section*{Significance statement}
%
%Geometric fluid \emph{mechanics} casts new light on the problem of
%\emph{Sargassum} inundation in the Caribbean Sea.  On one hand,
%recent nonlinear dynamical systems results pertaining to the fluid
%\emph{kinematics} identify the carriers of \emph{Sargassum} with
%coherent Lagrangian vortices whose (flow-invariant) boundaries defy
%stretching.  These vortices possess finite-time attractors for the
%cargo, viz., \emph{Sargassum} rafts modeled as elastic networks of
%inertial particles, which makes transportation by ocean currents
%and winds effective.  On the other hand, a two-dimensional model
%of baroclinic Caribbean Sea eddy \emph{dynamics} with buoyancy
%inhomogeneity and Lie--Poisson Hamiltonian structure identifies
%thermal instability mediated by bottom topography as a mechanism
%for filamentation and ensuing coastal inundation.  The results are
%consequential for the prediction of \emph{Sargassum} arrival, and
%thus for response and planning.

\section{Introduction}

Over this past decade, beaching events of pelagic \emph{Sargassum},
a type of brown macroalgae that forms floating rafts at the ocean
surface, have been reported nearly every spring and summer within
the Caribbean Sea \citep{Wang-etal-19}.  These rafts of algae serve
as habitats for marine fauna \citep{Bertola-etal-20} and can be an
important carbon sink with consequences for global climate regulation
\citep{Paraguay-etal-20}.  At the same time, they can carry high
levels of arsenic and heavy metals, producing major problems when
decomposing on beaches such as negatively impacting seagrass
communities, corals, and water quality with an increase in sea
turtle and fish mortality, causing health problems in humans,
diminishing tourism and, as a result, disrupting the local economy
\citep{Smetacek-Zingone-13, Resiere-etal-18}.

The negative consequences of \emph{Sargassum} choking in the coasts
of the Caribbean Sea requires improvement in forecasting of
\emph{Sargassum} beaching events to allow coastal zone managers and
decision makers to timely prepare and respond adequately.  This
work contributes, in part, to fulfill this demand by unveiling
especial ocean phenomena that have the ability to facilitate the
transport of \emph{Sargassum} that conduct to subsequent ``waves''
of coastal inundation.  It also seeks to get a basic insight into
the process leading to the latter.

\begin{figure}[h!]
  \centering%
  \includegraphics[width=\textwidth]{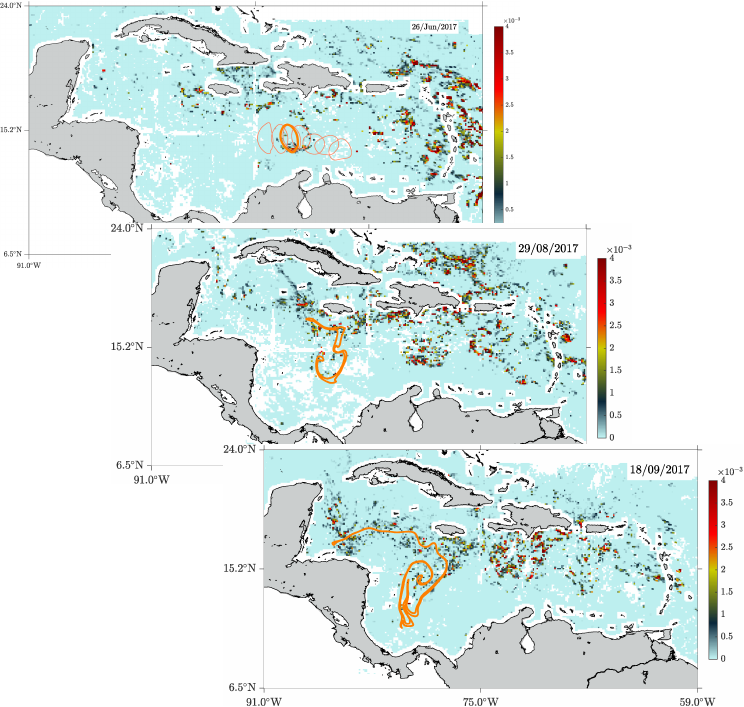}%
  \caption{Sequence of images showing satellite-derived \emph{Sargassum}
  distribution (percentage of coverage within a pixel) on the ocean
  surface in the Caribbean Sea. White represents absence of data.
  Overlaid on each image is the boundary of \emph{Kukulkan}, an
  anticyclonic mesoscale eddy detected from satellite altimetry
  that was possible to be classified as coherent in a Lagrangian
  (i.e., fluid following) sense for two months.  \textcolor{black}{Forward
  and backward trailing advected images of the vortex boundary
  during that period, ranging from 15/May/2017 to 14/July/2017 (cf.\
  Fig.\ \ref{fig:kukulkan}), are depicted in thin in the top panel.}}
  \label{fig:fad}%
\end{figure}

The sequence of images in Fig.\ \ref{fig:fad} corresponds to
satellite-inferred \emph{Sargassum} distribution on the surface of
the ocean in the Caribbean Sea as obtained from the 7-day Floating
Algae Density (FAD) \citep{Trinanes-etal-21}.  On any given day,
the FAD represents the average percentage of \emph{Sargassum}
coverage within a unit area or an image pixel over the last seven
days, ending on the given day. The FAD is computed from the Alternative
Floating Algae Index (AFAI), which serves as a measure of the
magnitude of MODIS (Moderate Resolution Imaging Spectroradiometer)
red edge reflectance of floating vegetation \citep{Wang-Hu-16}.
Overlaid in yellow on each FAD field is a snapshot of the material
(i.e., Lagrangian) boundary of an anticyclonic (i.e., clockwise
rotating) mesoscale eddy, which we have named \emph{Kukulkan}.
Detected from satellite-altimetry measurements of sea-surface height
(SSH) \citep{LeTraon-etal-98}, the boundary of \emph{Kukulkan} was
found to experience literally no stretching from 15/May/2017 to
14/July/2017.  However, despite the fact that the boundary of
\emph{Kukulkan} represents a barrier for fluid transport, on
26/Jun/2017 \emph{Sargassum} is seen to spiral inward from the
region surrounding \emph{Kukulkan} to its interior, bypassing its
boundary. While \emph{Kukulkan} drifts westward, it carries
\emph{Sargassum} within. Eventually, as it encounters in its path
shallower and shallower water, it destabilizes.  This process is
characterized by intense filamentation (reminiscent of a writhing
sneak, as is \emph{Kukulkan}---the Mesoamerican feathered serpent
deity---commonly depicted). The filaments breaking away from
\emph{Kukulkan}, and the \emph{Sargassum} carried within, reach the
continental margins of Central America and eventually also the
Yucatan Peninsula.

We dedicate the next section (\ref{sec:carriers}) to show that
eddies with persistent coherent material boundaries such as
\emph{Kukulkan}---\emph{the carriers} of \emph{Sargassum}--commonly
traverse the Caribbean Sea by building from incoherent fluid.  This
will be preceded by a review of the nonlinear dynamics technique
\citep{Haller-Beron-13, Haller-Beron-14} used to frame them. In
Section \ref{sec:cargo} we articulate how ``coherent Lagrangian
eddies'' have the capacity of capturing \emph{Sargassum} rafts---\emph{the
cargo}---and of dragging them along.  This will make use of recent
results \citep{Beron-21-ND} pertaining to the dynamics of finite-size
or ``inertial'' particle motion on the air--sea interface under the
combined action of ocean currents and winds.  In Section
\ref{sec:mechanism} we discuss \emph{the mechanism} that leads to
\emph{Sargassum} coastal inundation.  This is based on the proposition
of a minimal model for Caribbean Sea vortex dynamics, whose properties
are discussed in the Appendix.  The model builds on an old recipe,
which used to be very common in ocean dynamics \citep{Ripa-GAFD-93}
and is regaining momentum \citep{Kurganov-etal-20, Beron-21-RMF,
Beron-21-POFa, Beron-21-POFb, Holm-etal-21}, to include thermodynamics
in the two-dimensional rotating shallow-water model.  The paper is
closed with a summary and some concluding remarks in Section
\ref{sec:conclusions}.

\section{The carriers}\label{sec:carriers}

Let $\u(\x,t)$ be a two-dimensional fluid velocity, with $\x$
denoting position in some domain of $\mathbb R^2$ and $t$ referring
to time. Let $\varphi_{t_0}^t$ be the flow map associating
fluid particle positions at times $t_0$ and $t$, which follows by
integrating the motion equation, viz., $\dot\x = \u(\x,t)$.

The notion of a vortex with a material boundary resisting stretching
under advection by the flow, e.g., inferred geostrophically from
altimetry data, from time $t_0$ to time $t_0 + T$ for some (finite) $T$
is expressed by the variational principle \citep{Haller-Beron-13,
Haller-Beron-14}
\begin{equation}
  \delta\oint 
  \frac
  {\sqrt{\mathbf r'(s)\cdot C_{t_0}^{t_0+T}(\mathbf r(s))\mathbf r'(s)}}
  {\sqrt{\mathbf r'(s)\cdot \mathbf r'(s)}}\d{s}
  = 0.
  \label{eq:variational}
\end{equation}
Here, $\mathbf r(s)$ provides a parametrization for a material loop
at time $t_0$ and $C_{t_0}^{t} := (\D{\varphi_{t_0}^t})^\top
\D{\varphi_{t_0}^t}$, where $\D{}$ denotes derivative with respect
to time-$t_0$ position, is the (symmetric, positive-definite)
\emph{Cauchy--Green strain tensor field}.  The integrand in
\eqref{eq:variational}, which objectively (i.e., independent of the
observer's viewpoint) measures relative stretching from $t_0$ to
$t_0 + T$, is symmetric under $s$-shifts and thus by Noether's
theorem it must be equal to a positive constant, say $p$.  In
other words, solutions to \eqref{eq:variational} are characterized
by uniformly $p$-stretching loops.  The time-$t_0$ positions of
such \emph{$p$-loops} turn out to be limit cycles of one of the
following two bidirectional vector or \emph{line} fields:
\begin{equation}
  \mathbf l_p^\pm(\mathbf r) :=
  \sqrt{
  \frac
  {\lambda_2(\mathbf r) - p^2}
  {\lambda_2(\mathbf r) - \lambda_1(\mathbf r)}
  }
  \,\v_1(\mathbf r) 
  \pm
  \sqrt{
  \frac
  {p^2 - \lambda_1(\mathbf r)}
  {\lambda_2(\mathbf r) - \lambda_1(\mathbf r)}
  }
  \,\v_2(\mathbf r),  
  \label{eq:p-line}
\end{equation}
where $\lambda_1 < p^2 < \lambda_2$.  Here, $\{\lambda_i\}$ and
$\{\v_i\}$, satisfying $0 < \lambda_1 \le \lambda_2$, $\v_i\cdot
\v_j = \delta_{ij}$, $i,j = 1,2$, are eigenvalues and (orientationless)
normalized eigenvectors, respectively, of $\smash{C_{t_0}^{t_0+T}}$.
Limit cycles of \eqref{eq:p-line} either grow or shrink under changes
in $p$, forming smooth annular regions of nonintersecting loops.
The outermost member of such a band of material loops is observed
physically as the boundary of a \emph{coherent Lagrangian eddy},
namely, a \emph{Lagrangian coherent structure} \citep{Haller-15}
of elliptic type that generalizes the notion of KAM torus to the
finite-time-aperiodic flow case \citep{Haller-Beron-12}.  The
$p$-loops can also be interpreted as so-called null-\emph{geodesics}
of the (sign-indefinite) \emph{generalized Green--Lagrangian tensor
field}, $C_{t_0}^{t_0+T}-\lambda\Id$.

\begin{remark}\label{rem:lavd}
  Two observations that follow from numerical experimentation
  \citep{Andrade-etal-20} are in order.  First, the boundaries of
  coherent material vortices revealed from geodesic detection are
  not only resisting stretching, but also are nearly diffusion
  resisting \citep{Haller-etal-18}.  Second, a stretching (or
  diffusion) withstanding Lagrangian eddy over $[t,t']$ typically
  includes, at $t$, a local maximum of the \emph{Lagrangian-averaged
  vorticity deviation} (or \emph{LAVD}), defined by
  \citep{Haller-etal-16}
  \begin{equation}
	 \mathrm{LAVD}_t^{t'}(\x) := \smash{\int_t^{t'}} |\xi(
	 \varphi_t^\tau(\x),\tau) - \bar{\xi}(\tau)|\d{\tau},
	 \label{eq:LAVD}
  \end{equation}
  where $\xi(\x,t)$ is the (vertical) vorticity of the fluid and
  $\bar{\xi}(t)$ is its average over the tracked fluid bulk.
\end{remark}

\begin{figure}[h!]
  \centering%
  \includegraphics[width=\textwidth]{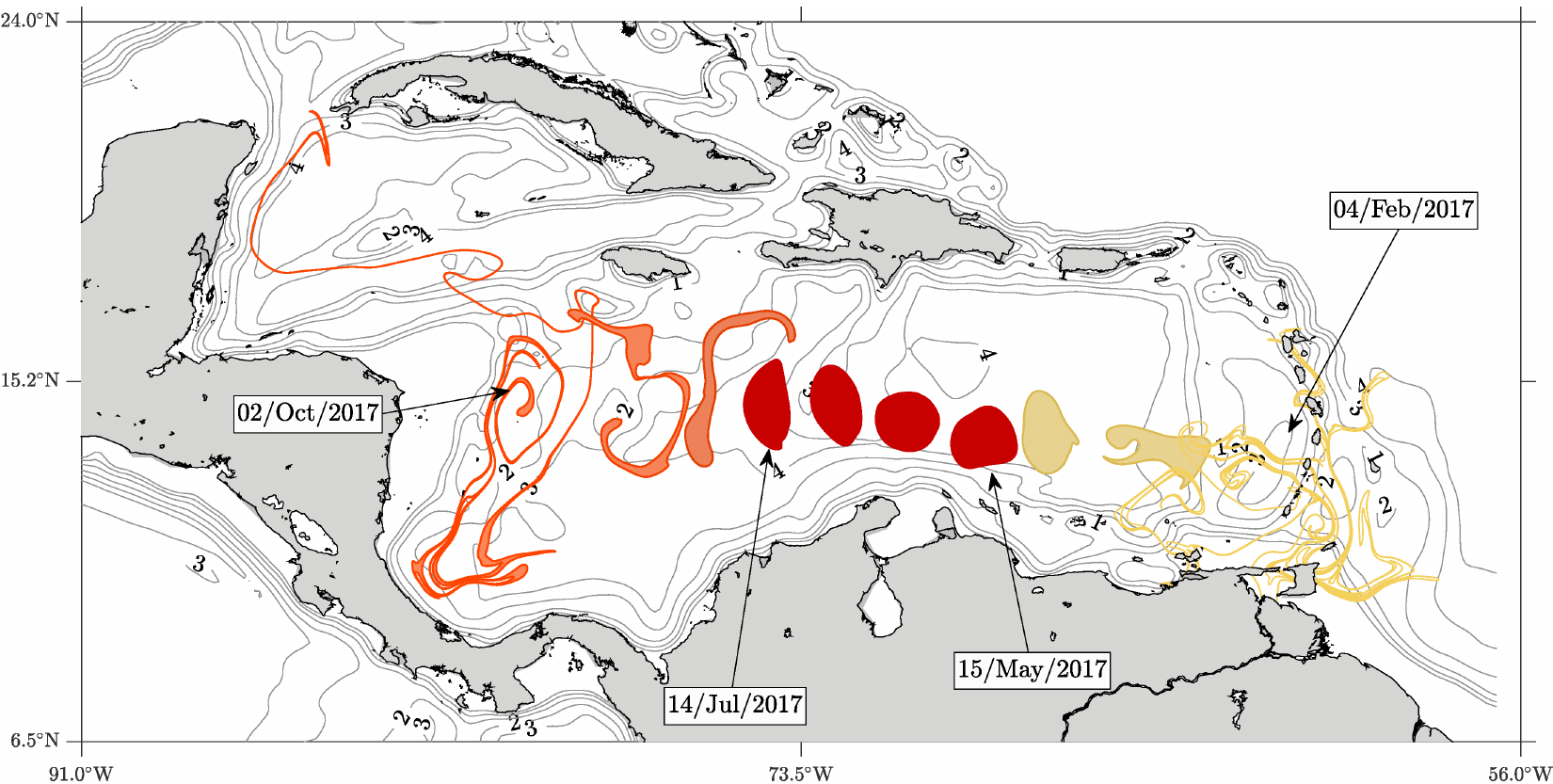}%
  \caption{\textcolor{black}{Genesis (yellow), evolution (red), and
  apocalypse (orange)} of \emph{Kukulkan}, an anticyclonic mesoscale
  vortex extracted from altimetry-derived surface geostrophic flow
  using geodesic detection. Selected isobaths (in km) are shown in
  gray.}
  \label{fig:kukulkan}%
\end{figure}

In Fig.\ \ref{fig:kukulkan} we illustrate the full evolution of
\emph{Kukulkan} since its genesis by using the methodology devised
by \citep{Andrade-etal-20}. This consists in repeatedly applying
\emph{geodesic eddy detection} on the (altimetric) flow domain of
definition, exhaustively searching the two-parameter space $(t_0,T)$.
\textcolor{black}{More precisely, we roll the initial time instance
$t_0$ over a time window covering the time interval of during which
a vortex is expected to exist.  For each $t_0$, we progress $T$ as
long as the Lagrangian method successfully detects a coherent vortex.
Thus, we obtain for each $t_0$ a life expectancy $T_{\max}(t_0)$,
which is the maximum $T$ for which a Lagrangian simulation starting
at $t_0$ successfully detected a coherent vortex.  The expected
results is a wedge-shaped $T_{\max}(t_0)$ distribution, indicating
that all Lagrangian coherence assessments predict the breakdown
consistently, independent of any parameter presets.  Robust assessments
of the birth and decease dates of the vortex are obtained by combining
the results from running the algorithm in forward- and backward-time
directions.}  The detection scheme is numerically implemented in
the Julia package
\href{https://github.com/CoherentStructures/CoherentStructures.jl}{\texttt{CoherentStructures.\allowbreak
jl}} by \citep{Karrasch-Schilling-20} using the index theory for
$p$-line fields \eqref{eq:p-line} developed in \citep{Karrasch-etal-14}.
As revealed in Fig.\ \ref{fig:kukulkan}, \emph{Kukulkan} builds
material coherence out of fully incoherent fluid (water) that
penetrates through the Lesser Antilles passages from the Atlantic
Ocean.  This happens on 15/May/2017.

\begin{remark}
  While it is beyond the scope of this paper to explain the process
  that leads to the birth of \emph{Kukulkan} and other eastern
  Caribbean Sea coherent Lagrangian eddies, this is certainly much
  more involved than the result of the squeezing of lens-like eddies
  (North Brazil Current rings) through gaps (Lesser Antilles passages)
  as envisioned by \citep{Simmons-Nof-02}.  Moreover, while eastern
  Caribbean Sea eddies can carry traces of Amazon and Orinoco Rivers
  water properties and sediments \citep{vanderBoog-etal-19}, recent
  claims \citep{Huang-etal-21} on the role of ``SSH eddies'' in
  connecting the tropical Atlantic Ocean and the Gulf of Mexico are
  unfounded.  Identified as regions instantaneously encircled by
  SSH level curves, SSH eddies do not possess material boundaries.
  Thus, they cannot hold and carry within fluid long distances as
  coherent Lagrangian eddies to make connectivity assessments.  The
  two main issues with this type of Eulerian eddy detection, which
  is the de-facto detection method in oceanography, are 1) its lack
  of objectivity, which leads to many false positives and also
  negatives \citep{Beron-etal-15}, and, as discussed, 2) its lack
  of flow invariance.
\end{remark}

Since its birth date, namely, the day in which it is first extracted
from altimetry using geodesic detection, \emph{Kukulkan} translates
westward preserving its material coherence for nearly three months,
in the rigorous sense that its boundary stretches by a factor $p =
1.192$. (We use red color in Fig.\ \ref{fig:kukulkan} to depict the
vortex while it is classified as Lagrangian coherent.)  Eventually,
starting on 14/July/2017, by the mechanism proposed below,
\emph{Kukulkan} loses its material coherence, spreading its contents,
mainly over the continental margins of Central America and the
Yucatan Peninsula, along long filaments that break away from the
vortex.

\begin{figure}[h!]
  \centering%
  \includegraphics[width=\textwidth]{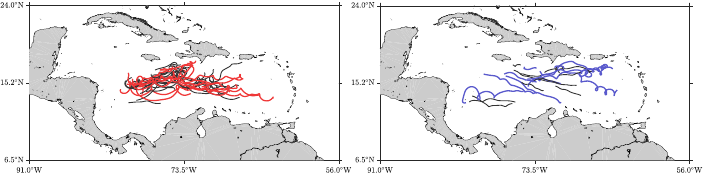}%
  \caption{Trajectories of coherent Lagrangian eddies detected
  geodesically from altimetry over 2011--2019. Anticyclones (resp.,
  cyclones) are depicted in the left (resp., right) panel. A gray
  trajectory indicates that the vortex was not seen to carry
  \emph{Sargassum} or that the satellite imagery data were not
  sufficient to conclude it did.}
  \label{fig:census}%
\end{figure}

Since 2011, the year when the first major \emph{Sargassum} event
was recorded, there have been many other vortices traversing the
eastern Caribbean Sea with a degree of material coherence similar
to \emph{Kukulkan}.  We depict in Fig.\ \ref{fig:census} the
trajectories of geodesically detected vortices from satellite
altimetry over 2011--2019.  Anticyclones (left panel) are as frequent
as cyclones (right panel), which are smaller, have shorter lifetimes,
and stretch more than anticyclones (Table \ref{tab:census}).

\renewcommand{\arraystretch}{1.25}
\begin{table}
  \centering
  \begin{tabular}{lcccc}
	 \hline\hline
	 Polarity & No.\ per year & $T$ [d] & $p$ & Mean radius [km]\\
	 Cyclonic & $2 \pm 2$ & $52 \pm 34$ & $1.23 \pm 0.33$ & $76 \pm 20$\\
	 Anticyclonic & $4\pm3$ & $68\pm40$ & $1.23\pm0.34$ & $81\pm32$\\
	  \hline
  \end{tabular}
  \caption{Characteristic parameters of coherent Lagrangian eddies
  detected geodesically from altimetry over 2011--2019. Stated values
  are mean values, each one accompanied by a one-standard-deviation
  uncertainty.} \label{tab:census}
\end{table}
\renewcommand{\arraystretch}{1}

\section{The cargo}\label{sec:cargo}

The notion that mesoscale vortices with material boundaries such
as \emph{Kukulkan} represent effective carriers of \emph{Sargassum},
and, thus, through filamentation contribute to coastal inundation,
is supported on the dynamics of elastically connected networks of
finite-size buoyant or ``inertial'' particles \citep{Beron-Miron-20},
which are especial near so-called rotationally coherent vortices,
i.e., with material boundaries given by convex level curves of the
LAVD field \eqref{eq:LAVD} surrounding LAVD maxima \citep{Haller-etal-16}.
Pelagic \emph{Sargassum} rafts are composed of flexible stems kept
afloat by gas-filled bladders.  Elastic networks of inertial particles
provide a minimal representation for them.  Building on the
Maxey--Riley equation for the dynamics of inertial particles floating
at the air--sea surface of \citep{Beron-etal-19-PoF}, referred to
as the BOM equation, \citep{Beron-Miron-20} show that the motion
of an elastic network with $N$ inertial particles obeys
\begin{subequations}
\begin{equation}
  \ddot \x_i + \left(\left. f\right\vert_i + \tfrac{1}{3}\varrho\left.\xi
  \right\vert_i\right)\dot \x_i^\perp + \frac{\dot \x_i}{\tau} =
  \varrho\dot\u\vert_i + \varrho\left(\left. f\right\vert_i +
  \tfrac{1}{3}\left.\xi\right\vert_i\right) \u\vert_i^\perp +
  \frac{(1-\alpha)\u\vert_i + \alpha \mathbf
  w\vert_i}{\tau} + \mathbf F_i, 
  \label{eq:mr}
\end{equation}
$i = 1,\dotsc,N$, where
\begin{equation}
   \mathbf F_i = - \sum_{j\in \neigh(i)} k_{ij}\big(|\x_i - \x_j| -
   \ell_{ij}\big)\frac{\x_i - \x_j}{|\x_i - \x_j|},
   \label{eq:F}
\end{equation}
\label{eq:MR}%
\end{subequations}
and $\left. \right\vert_i$ means pertaining to particle $i$.  The
system of coupled ordinary differential equations \eqref{eq:MR}
represents a Newton law which includes, in addition to the forces
included in the BOM equation (flow, added mass, lift, drag, and
Coriolis), the elastic force (Hook's law) exerted on each particle
by neighboring particles \eqref{eq:F}, where $k_{ij} > 0$ is the
stiffness of the spring connecting particles $i$ and $j$ and
$\ell_{ij} > 0$ is the length of the latter at rest.  The rest of
the variables and parameters in \eqref{eq:mr} is as follows: $\x_i(t)$
is the instantaneous position of particle $i$; $\u(\x,t)$ is the
\emph{near-surface} ocean velocity, with $\dot\u(\x,t)$ denoting
its total (material) derivative and $\xi(\x,t)$ its (vertical)
vorticity; $\w(\x,t)$ is near-surface wind velocity; $f$ is the
Coriolis parameter; and the parameters $\varrho \in [0,1)$, $\tau
> 0$ and $\alpha \in [0,1)$ depend on the water-to-particle density
ratio $\delta \ge 1$ in specific forms as given in
\citep{Beron-etal-19-PoF} \citep[cf.\ also][]{Olascoaga-etal-20,
Miron-etal-20-GRL, Miron-etal-20-POF}.  In particular, $\tau$
measures the inertial response time of the medium to the particle,
and is proportional to the square of the radius of the latter. In
turn, $\alpha$ can be interpreted as a buoyancy-dependent leeway
factor, terminology commonly used in the search-and-rescue-at-sea
literature \citep{Breivik-etal-13}.  An informal statement of Theorem
4.1 of \citep{Beron-Miron-20} is as follows:
\begin{quote}
  \emph{the trajectory of the center of a rotationally coherent
  vortex that spins anticyclonically (resp., cyclonically) is locally
  forward attracting over the lifetime of the vortex for all $k_{ij}$
  (resp., if $\sum_{i=1}^N\sum_{j\in\neigh(i)} k_{ij}$ is larger
  than a quantity that decays with $N$ to a typically small value)
  provided that winds are sufficiently calm.}
\end{quote}
Moreover, numerical experimentation reveals that the above statement
holds under fairly general wind conditions \citep{Beron-Miron-20}.

By view of the second observation in Remark \ref{rem:lavd}, coherent
Lagrangian vortices detected geodesically from altimetry-derived
velocity should attract \emph{Sargassum} rafts.  This is consistent
with \emph{Kukulkan} collecting \emph{Sargassum}, transporting it,
and eventually, upon destabilization, spreading the \emph{Sargassum}
in the surrounding areas, which include the continental margins of
the Yucatan Peninsula and Central America.

Figure \ref{fig:census} reveals that the behavior predicted by the
theory of \citep{Beron-Miron-20} applies to a large number of the
coherent Lagrangian eddies travelling through the eastern Caribbean
Sea.  The trajectories of such vortices are indicated in color; the
black lines correspond to trajectories of eddies that either did
not capture \emph{Sargassum} or the quality of the satellite imagery
data was too poor (e.g., due to excessive cloud coverage) to determine
that they did.  More precisely, out of a total of 40 (resp., 21)
geodesically detected anticyclones (resp., cyclones) over 2011--2019,
17 (resp., 10) were visually found to have \emph{Sargassum} spiraling
inward at some instant along their trajectories.  

\section{The mechanism}\label{sec:mechanism}

While in-situ sampling is not exhaustive, a recent hydrographic
survey by \citep{vanderBoog-etal-19} suggests that mesoscale
anticyclones in the eastern Caribbean Sea are surface intensified.
This is puzzling inasmuch as surface-intensified vortices should
be able to relatively easily bypass topographic obstacles according
to linear theory arguments and fully nonlinear numerical simulations
by \citep{Adams-Flierl-10}. This assessment, however, was based on
the consideration of a model with two homogeneous density layers.
We explore here how this scenario changes by making such a model
more realistic through the incorporation of \emph{density inhomogeneity
effects}.  This is done in the simplest manner possible, by allowing
the density (temperature) in the upper layer to vary laterally and
with time, while keeping it as well as the (horizontal) velocity
as depth independent \citep[e.g.,][]{Ripa-GAFD-93}.

Our minimal two-layer model is formulated as follows.  The upper
layer has, as anticipated, density varying in lateral position, $\x
= (x,y)$, and time, $t$, and is limited from above by a horizontal
rigid lid.  Position $\x$ is assumed to range on a domain $D$ of
the $\beta$ plane, so $x$ (resp., $y$) points eastward (resp.,
northward).  Specifically, $D = [0,30R] \times [0,10R]$, centered
in the eastern Caribbean Sea, where
\begin{equation}
  R^2 := \frac{g'H/f_0^2}{1+r^{-1}}
\end{equation}
is the square of the Rossby radius of deformation. Here, $f_0 > 0$
is the mean Coriolis parameter in $D$; $g' > 0$ is the
buoyancy\footnote{The \emph{buoyancy} is gravity times the ratio
of the difference between the lower and upper layer densities to
the (constant) Boussinesq approximation's density.} jump across the
interface between the upper and lower layers in the reference state,
i.e., with constant density in the upper layer and no currents in
either layer; $H > 0$ is the mean thickness of the lower layer; and
$rH$, $r>0$, is the mean thickness of the upper layer.  From the
global climatological data \citep{Chelton-etal-98} it follows that
$R = 60$ km is representative of eastern Caribbean Sea conditions.
This justifies our domain size choice.  From \citep{vanderBoog-etal-19}
in-situ hydrographic observations, $r = 0.2$ seems fair (the depth
in the eastern Caribbean Sea can reach 5 km or more; cf.\ Fig.\
\ref{fig:kukulkan}).  The lower layer has homogeneous density, is
(much) heavier than the upper layer, and rests on a rigid bottom
with at most a constant zonal slope $-H\beta_\mathrm{T}/f_0$, where
$\beta_\mathrm{T} > 0$ is a constant.  Assuming quasigeostrophic
(QG) dynamics, the model equations take the form
\begin{subequations}
\begin{equation}
  \left.
  \begin{aligned}
	 \partial_t\xx_i + [\p_i,\xx_i] - \delta_{1i} R^{-2}[\p_1,\ps]
	 &= 0,\\
    \partial_t\ps + [\p_1,\ps] &= 0,
  \end{aligned}
  \right\}
\end{equation}
$i=1,2$.  Here, $[A,B] := \nabla^\bot A \cdot \nabla B$ is the
Jacobian for (time-dependent) functions $A$ and $B$ on $\mathbb
R^2$, and the upper ($i=1$) and lower ($i=2$) layer streamfunctions
$(\p_1(\x,t), \p_2(\x,t))$ are related to the fields ($\xx_1(\x,t),
\ps(\x,t), \xx_2(\x,t))$ through the invertibility principle
\begin{equation}
  \nabla^2\p_i +
  \frac{(-1)^i R^{-2}}{1+r^{(-1)^{i+1}}}\big(\p_1-\p_2\big) =
  \xx_i - \delta_{1i}R^{-2}\ps - \beta y -
  \delta_{2i}\big(1-\theta(x-15R)\big)\beta_\mathrm{T} x,
  \label{eq:invertibility}
\end{equation}
\label{eq:IL0HL}%
\end{subequations}
where $\theta$ is the Heaviside step function. Note that the
topographic slope is nonzero eastward of the mid zonal point of the
domain, minimally representing the actual topography in the eastern
Caribbean Sea.  This is roughly achieved by setting $\beta_\mathrm{T}
= 0.005f_0/H$.  The field $q_i$ represents the $i$th-layer QG
potential vorticity and $2g'\ps/f_0R^2$ is the QG buoyancy deviation
from the reference buoyancy ($g'$) in the upper layer. The subspace
$\{\ps = \const\}$ is invariant.  The dynamics on it are governed
by the QG model with two homogeneous layers \citep[e.g.,][]{Ripa-JFM-91}.
Setting $\psi_2 = 0$ and making $H \to \infty$ while keeping $rH$
finite, leads to the inhomogeneous-layer reduced-gravity QG model
originally developed in \citep{Ripa-RMF-96}.  See \citep{Beron-21-POFa,
Beron-21-POFb, Holm-etal-21} for recent discussions on geometric
aspects of this model as well as on sustained ``thermal'' instabilities
\citep{Gouzien-etal-17}, and \citep{Crisan-etal-21} for the
construction of unique solutions.  Well-posedness of \eqref{eq:IL0HL}
is here supported on its uniqueness of solutions and manifest
generalized (noncanonical) Hamiltonian structure (cf.\ Appendix).

The simulations are initialized by specifying $\psi_i$ and $\psi$.
Specifically, we set 
\begin{equation}
  \psi_i(\x,0) = \sqrt{\mathrm{e}} R V_i \mathrm{e}^{\frac{-(x -
  27R)^2 - (y - 7R)^2}{2R^2}},
  \label{eq:initial}
\end{equation}
with $V_1 = 0.75$ m\,s$^{-1}$, supported on \citep{vanderBoog-etal-19}
in-situ hydrographic measurements.  This represents an initially
localized vortex-like structure with anticyclonic polarity.  Without
observational support we choose $V_2 = V_1/4$, yet consistent with
earlier characterizations of surface-intensified vortices
\citep{Adams-Flierl-10}, and $\psi(\x,0) = \psi_2(\x,0)$.  We also
impose a background westward flow in upper layer of $0.1$ m\,s$^{-1}$,
estimated from the time coherent Lagrangian eddies detected from
altimetry typically take to traverse the eastern Caribbean Sea.
This is on the order of  the mean zonal velocity of the Caribbean
Current based on satellite-tracked surface drifter trajectories
\citep{Richardson-05}.  We use the same background westward velocity
in the lower layer to minimize the development of baroclinic
instability.  This choice is not an artifact, but is actually
supported by the observation, presented in the preceding section,
that eastern Caribbean Sea eddies can remain materially coherent
for months. The results are not sensitive to the other parameter
choices.

\begin{figure}[h!]
  \centering%
  \includegraphics[width=\textwidth]{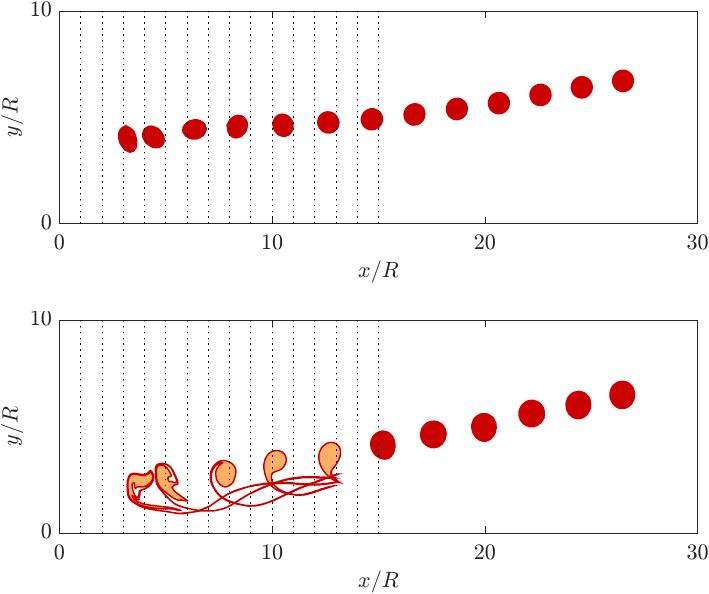}% 
  \caption{Snapshots reflecting the evolution every 15 d of a
  surface-intensified Lagrangian vortex classified as coherent by
  geodesic detection over the period that the structure is depicted
  in red, according to the quasigeostrophic two-layer model
  \eqref{eq:IL0HL} with homogeneous (top) and inhomogeneous (bottom)
  density in the upper layer.  Broken lines indicate isobaths, with
  depth decreasing linearly to the west (left). Length is scaled
  by the (internal) Rossby deformation radius.}
  \label{fig:mechanism}%
\end{figure}

The results strongly depend, however, on whether density inhomogeneity
(in the upper layer) is allowed or not.  This is illustrated in
Fig.\ \ref{fig:mechanism}, with each panel showing snapshots, every
$15$ d, of the westward translation of a structure revealed as a
coherent Lagrangian vortex using geodesic detection.  The upper
(resp., lower) panel corresponds to a simulation with homogeneous
(resp., inhomogeneous) density.  The simulations were performed on
$D = \mathbb R/30R\mathbb Z \times \mathbb R/10R\mathbb Z$, for
simplicity as commonly done \citep{Adams-Flierl-10}, using a
fully-dealiased pseudospectral code on a $512 \times 256$ grid with
a fourth-order Runge--Kutta time stepper for $(q_1 - \beta y, \psi,
q_2 - \beta y - \beta_\mathrm{T}x)$ and bi-Laplacian hyperviscosity
\citep[cf.][for details]{Beron-etal-08-JAS}.  In each panel the
first red blob to the right (east) corresponds to the vortex at the
instance it first acquires Lagrangian coherence.  This happens
approximately 5 d after initialization at $\x=(27R,7R)$ for the
simulation without inhomogeneous density and with it.  Both the
homogeneous and inhomogeneous vortex take a southwestward trajectory,
consistent with numerical simulations and rotating-tank experiments
\citep{Carnevale-etal-91b} and altimetry observations
\citep{Morrow-etal-04a}.  We depict in red the advected images of
each vortex while classified as Lagrangian coherent, and in orange
the advected images past the Lagrangian coherence horizon.  The
homogeneous vortex remains coherent, in the most strict geodesic
sense with stretching factor $p \approx 1$, for the extent of the
simulation, 145 d, while the inhomogeneous vortex for only 90 d.
The condition $p \approx 1$ and area preservation (which follows
from $\nabla\cdot\nabla^\bot\psi_1 = 0$) impose a strong restriction
on the deformation of these vortices respectively over $t \in
[5,150]$ d and $t \in [5,95]$ d, as is evident from Fig.\
\ref{fig:mechanism}.  The homogeneous vortex is much more longevous
than the inhomogeneous one.  It bypasses the topographic slope,
consistent with earlier predictions \citep{Adams-Flierl-10}.  The
only evident effect on the vortex is a slight change in its trajectory.
In stark contrast, the inhomogeneous vortex experiences vigorous
filamentation as soon as it starts to travel over shallower fluid
(the vertical broken isolines indicate isobaths).

Our minimal model of surface-intensified Caribbean Sea vortex
dynamics thus identifies thermal instability in the presence of
bottom topography as a mechanism for vortex filamentation.  This is
consequential for \emph{Sargassum} coastal inundation as articulated
in the previous sections.

\begin{remark}
 Self-induced westward propagation \citep{Nof-81a} is hard to realize
 in the simulations.  On the other hand, if these are initialized
 with a cyclonic vortex structure, the trajectories result
 northwestward, consistent with the numerical, experimental, and
 observational evidence given in \citep{Carnevale-etal-91b,
 Morrow-etal-04a}.  The eastern Caribbean Sea, by contrast with our
 minimal model simulations, seems too constrained in longitude for
 coherent Lagrangian vortices detected from altimetry to clearly
 reveal a preferred meridional direction in their westward drift.
 Only cyclones reveal a predominantly northwestward translation
 pattern (cf.\ Fig.\ \ref{fig:census}).  The inhomogeneous vortex
 takes 125 d to cover the same longitudinal distance that the
 homogeneous vortex covers in 145 d.  Inhomogeneity thus contributes
 to speeding up the vortex translation.  Finally, that material
 coherence can be realized for the noted extended periods of time,
 particularly in the homogeneous case, is surprising inasmuch as
 \eqref{eq:initial} does not represent a solution of \eqref{eq:IL0HL}
 and Rossby-wave radiation invariably happens immediately after
 initialization \citep[e.g.][]{Adams-Flierl-10}.
\end{remark}

\section{Summary and concluding remarks}\label{sec:conclusions}

We have shown that \emph{Sargassum} transport in the eastern Caribbean
Sea can be effectively accomplished by coherent Lagrangian mesoscale
eddies.  Geodesically detected from satellite altimetry, these
vortices, despite having material boundaries that resist stretching
and are impassable by fluid particles, contain finite-time attractors
for \emph{Sargassum}.  This was confirmed by satellite observations
in agreement with the prediction by the Maxey--Riley theory for the
motion of elastic networks of finite-size particles floating at the
ocean surface.  The coastal inundation of \emph{Sargassum} was found
to be associated with the filamentation that the vortices experience
as they encounter shallower water in their westward drift.  Thermal
instability of surface-intensified vortices interacting with
topography was identified as a mechanism leading to the filamentation.
This was done by proposing a minimal model for (eastern) Caribbean
Sea vortex dynamics with solid geometric properties that revisits
a simple recipe, once very popular, then abandoned, and now
experiencing a strong comeback, to incorporate thermodynamics in a
shallow-water model.  Our results are consequential for the prediction
of \emph{Sargassum} inundation events, and thus for planning an
adequate response to them.  The basic take-home message is:
\begin{quote}
  \emph{to aid in the forecasting of \emph{Sargassum} coastal
  arrivals, monitor the progression of mesoscale eddies geodesically
  detected from altimetry}.
\end{quote}
This might seem an impossible task as it appears to require unavailable
future altimetry-derived flow information. However, geodesic eddy
detection can be applied in \emph{backward time}, as shown in
\citep{Andrade-etal-20}.  In other words, material coherence
assessments can be made with observed velocity up to the assessment
instant.  Finally, from a theoretical view point, substantial work,
beyond the scope of the analysis tools used here, needs to be done
to understand the process that leads to the formation of coherent
material vortices, particularly eastern Caribbean Sea vortices, out
of incoherent fluid. Thus far this remains a mystery.

\section*{Acknowledgements}

This work was supported by the University of Miami's Cooperative
Institute for Marine \& Atmospheric Studies.

\section*{Author declarations}

\subsection*{Conflict of interest}

The authors have no conflicts to disclose.

\subsection*{Data availability}

This paper does not involve the use of data.  The floating algae
density fields are produced by USF and distributed by SaWS
(\href{https://optics.marine.usf.edu/projects/saws.html}{https://optics.marine.usf.edu/projects/saws.html}).
The altimeter products are produced by SSALTO/DUCAS and distributed
by AVISO with support from CNES
(\href{http://www.aviso.oceanobs}{http://www.aviso.oceanobs}).

\appendix

\section{Uniqueness of solutions and Hamiltonian structure of \eqref{eq:IL0HL}}

Let $\varphi_i := \xx_i - \delta_{1i}R^{-2}\ps - \beta y -
\delta_{2i}\big(1-\theta(x-10R)\big)\beta_\mathrm{T} x$.
Writing $\psi_i = (-1)^ir^{i-2}\hat\psi_1 + \hat\psi_2$ and
similarly for $\varphi_i$, from \eqref{eq:invertibility} one
finds
\begin{equation}
  \nabla^2\hat\psi_i - \delta_{1i}R^{-2}\hat\psi_i =
  \hat\varphi_i.
  \label{eq:uniqueness}
\end{equation}
Assume that $\hat\psi_i^{(1)}$ and $\hat\psi_i^{(2)}$ satisfy
\eqref{eq:uniqueness} for $i=1,2$ on $D \subseteq \mathbb R^2$ under
appropriate boundary conditions.  These are: 1)
$\smash{\nabla\hat\psi_i^{(j)}\times\n\vert_{\partial D} = 0}$ where
$\n$ is the unit normal to the solid boundary of $D$; 2)
$\smash{\hat\psi_i^{(j)}} \to 0$ as $\x\to\infty$ if $D$ spans
$\mathbb R^2$, or 3) $\smash{\hat\psi_i^{(j)}}(x+L_x) =
\smash{\hat\psi_i^{(j)}}$ and $\smash{\hat\psi_i^{(j)}}(y+L_y) =
\smash{\hat\psi_i^{(j)}}$ when $D = \mathbb R/L_x\mathbb Z\times
\mathbb R/L_y\mathbb Z$.  Eliminating $\hat\varphi_j$ from
\begin{equation}
  \nabla^2\hat\psi_i^{(j)} - \delta_{1i}R^{-2}\hat\psi_i^{(j)} =
  \hat\varphi_i,
\end{equation}
multiplying the result by $\smash{\psi_i^{(1)} - \psi_i^{(2)}}$,
and integrating by parts,
\begin{equation}
  \int_D\big|\nabla\big(\psi_i^{(1)} - \psi_i^{(2)}\big)\big|^2 +
  \delta_{1i}R^{-2}\big(\psi_i^{(1)} - \psi_i^{(2)}\big)^2 \d{x}\d{y}=
  0,
\end{equation}
from which it follows that $\hat\psi_i^{(1)} =\hat\psi_i^{(2)}$.
Thus if a solution to \eqref{eq:IL0HL} exists, then it is unique.

Now let $\U[\mu] = \int_DU(\x,\mu,\partial_x\mu, \partial_y\mu,
\partial_{xy}\mu, \dotsc)\d{x}\d{y}$ be a functional of sufficiently
smooth fields $\mu(\x) = (\mu^1(\x),\mu^2(\x),\dotsc)$ on $D$,
which will be assumed to either be bounded by a solid boundary or
span $\mathbb R^2$. The functional derivative of $\U$ with respect
to $\mu^a$, denoted $\smash{\del{\U}{\mu^a}}$, is the unique element
satisfying $\U[\mu^a+\varepsilon\delta\mu^a] - \U[\mu^a] =
\varepsilon\int_D \smash{\del{\U}{\mu^a}}\delta\mu^a\d{x}\d{y} +
O(\varepsilon^2)$ as $\varepsilon\to 0$.  We will say that $\U[\mu]$
is \emph{admissible} if $\nabla\smash{\del{\U}{\mu^a}}\cdot
\n\vert_{\partial D} = 0$ or $\smash{\del{\U}{\mu^a}} \to 0$ as
$\x\to\infty$. The set of admissible functionals, denoted $\mathcal
A$, cannot be extended to functionals of functions on $D = \mathbb
R/L_x\mathbb Z\times \mathbb R/L_y\mathbb Z$. With this in mind,
\eqref{eq:IL0HL} can be cast as a generalized (noncanonical)
Hamiltonian system \citep{Morrison-98}
\begin{equation}
  \partial_t\mu = \{\mu,\H\}
\end{equation}
for $\mu := (\xx_1,\ps,\xx_2)$, with Hamiltonian given by
\begin{equation}
  \H[\mu] := \frac{1}{2}\int_D
  H_1|\nabla\psi_1|^2 +
  H_2|\nabla\psi_2|^2 +
  \frac{f_0^2}{g'}(\p_1-\p_2)^2 \d{x}\d{y}
\end{equation}
and Lie--Poisson bracket \citep{Thiffeault-Morrison-00}
\begin{equation}
  \{\U,\V\}[\mu] := W^{ab}_c \int_D
  \mu^c\left[\del{\U}{\mu^a} \del{\V}{\mu^b}\right] \d{x}\d{y}
  \label{eq:PB}
\end{equation}
for all $\U,\V[\mu] \in \mathcal A$,\footnote{We are implicitly
assuming that $\mathcal A$ is \emph{closed}, namely, if $\U,\V[\mu]
\in \mathcal A$, then $\{\U,\V\}[\mu] \in \mathcal A$.} where the
(2,1)-tensor $W$ has coefficients $W^{11}_1 = W^{12}_2 = W^{21}_2
= (rH)^{-1}$, $W^{33}_3 = H^{-1}$, and zero otherwise. That
\eqref{eq:PB} represents a genuine bracket follows from $W^{ab}_c
= W^{ba}_c$ and $W^{ab}_cW^{a'b'}_a = W^{ab'}_cW^{b'a'}_a$, which
respectively imply antisymmetry for the bracket ($\{\U,\V\} =
-\{\V,\U\}$ for $\U,\V[\mu] \in \mathcal A$) and guarantee that it
satisfies the Jacobi identity ($\{\{\U,\V\},\W\} + \{\{\W,\U\},\V\}
+ \{\{\V,\W\},\U\} = 0$ for $\U,\V,\W[\mu] \in \mathcal A$).  The
associated infinite-family of invariant Casimirs is given by $\C[\mu]
= \int \xx_1C_1(\ps) + C_2(\ps) + C_3(\xx_2)$ for any $C_i$, which
commutes in the bracket with any $\U[\mu]\in \mathcal A$.

\begin{remark}
The geometry of \eqref{eq:PB} is most easily grasped by writing
it explicitly as
\begin{equation}
  \{\U,\V\} = \int_D 
  \frac{\xx_1}{rH}\left[\del{\U}{\xx_1},
  \del{\V}{\xx_1}\right] + 
  \frac{\ps}{rH}\left(\left[\del{\U}{\xx_1},
  \del{\V}{\ps}\right] - \left[\del{\V}{\xx_1},
  \del{\U}{\ps}\right]\right) + 
  \frac{\xx_2}{H}\left[\del{\U}{\xx_2},
  \del{\V}{\xx_2}\right] \d{x}\d{y}.
\end{equation}
Let $\mathfrak a$ be the \emph{Lie enveloping algebra} of
$\mathrm{SDiff}(D)$, the group of area preserving diffeomorphisms
in $D$.  The corresponding vector space is that of smooth time-dependent
functions in $D$, denoted $\mathcal F(D)$, and the Lie bracket is
given by the canonical Poisson bracket, $[\,,\hspace{.1em}]$.  The
Lie--Poisson bracket \eqref{eq:PB} represents a product for a
\emph{realization of a Lie enveloping algebra} on functionals in
the dual (with respect to the $\mathrm L_2$ inner product) of
$\mathfrak a \times \mathfrak a_\mathrm{s}$, where $\mathfrak
a_\mathrm{s}$ is the extension of $\mathfrak a$ by semidirect sum
to the vector space $\mathfrak a \times \mathcal F(D)$, with the
representation of $\mathfrak a$ on $\mathcal F(D)$ given by
$[\,,\hspace{.1em}]$ \citep[cf.][for details]{Thiffeault-Morrison-00}.
\end{remark}

\bibliographystyle{mybst-square} 
\bibliography{fot}

\begin{thebibliography}{47}
\expandafter\ifx\csname natexlab\endcsname\relax\def\natexlab#1{#1}\fi

\bibitem[Adams and Flierl(2010)]{Adams-Flierl-10}
{\rm Adams, D.~K. and Flierl, G.~R.} [2010].  {Modeled interactions of
  mesoscale eddies with the East Pacific Rise: Implications for larval
  dispersal}. {\em Deep Sea Research Part I: Oceanographic Research Papers\/}
  57, 1163--1176.

\bibitem[Andrade-Canto {\rm et~al.}(2020)Andrade-Canto, Karrasch and
  Beron-Vera]{Andrade-etal-20}
{\rm Andrade-Canto, F., Karrasch, D. and Beron-Vera, F.~J.} [2020].  {Genesis,
  evolution, and apocalyse of Loop Current rings}. {\em Phys. Fluids\/} 32,
  116603.

\bibitem[Beron-Vera(2021{\natexlab{{\rm a}}})]{Beron-21-POFb}
{\rm Beron-Vera, F.~J.} [2021{\natexlab{{\rm a}}}].  Extended shallow-water
  theories with thermodynamics and geometry. {\em Phys. Fluids\/} 33, 106605.

\bibitem[Beron-Vera(2021{\natexlab{{\rm b}}})]{Beron-21-RMF}
{\rm Beron-Vera, F.~J.} [2021{\natexlab{{\rm b}}}].  Multilayer shallow-water
  model with stratification and shear. {\em Rev. Mex. Fis.\/} 67, 351--364.

\bibitem[Beron-Vera(2021{\natexlab{{\rm c}}})]{Beron-21-ND}
{\rm Beron-Vera, F.~J.} [2021{\natexlab{{\rm c}}}].  {Nonlinear dynamics of
  inertial particles in the ocean: From drifters and floats to marine debris
  and \emph{Sargassum}}. {\em Nonlinear Dyn.\/} 103, 1--26.

\bibitem[Beron-Vera(2021{\natexlab{{\rm d}}})]{Beron-21-POFa}
{\rm Beron-Vera, F.~J.} [2021{\natexlab{{\rm d}}}].  Nonlinear saturation of
  thermal instabilities. {\em Phys. Fluid\/} 33, 036608.

\bibitem[Beron-Vera {\rm et~al.}(2008)Beron-Vera, Brown, Olascoaga, Rypina,
  Kocak and Udovydchenkov]{Beron-etal-08-JAS}
{\rm Beron-Vera, F.~J., Brown, M.~G., Olascoaga, M.~J., Rypina, I.~I., Kocak,
  H. and Udovydchenkov, I.~A.} [2008].  Zonal jets as transport barriers in
  planetary atmospheres. {\em J. Atmos. Sci.\/} 65, 3316--3326.

\bibitem[Beron-Vera and Miron(2020)]{Beron-Miron-20}
{\rm Beron-Vera, F.~J. and Miron, P.} [2020].  A minimal {M}axey--{R}iley model
  for the drift of \emph{{S}argassum} rafts. {\em J. Fluid Mech.\/} 904, A8.

\bibitem[Beron-Vera {\rm et~al.}(2015)Beron-Vera, Olascoaga, Haller, Farazmand,
  {Tri\~nanes} and Wang]{Beron-etal-15}
{\rm Beron-Vera, F.~J., Olascoaga, M.~J., Haller, G., Farazmand, M.,
  {Tri\~nanes}, J. and Wang, Y.} [2015].  {Dissipative inertial transport
  patterns near coherent Lagrangian eddies in the ocean}. {\em Chaos\/} 25,
  087412.

\bibitem[Beron-Vera {\rm et~al.}(2019)Beron-Vera, Olascoaga and
  Miron]{Beron-etal-19-PoF}
{\rm Beron-Vera, F.~J., Olascoaga, M.~J. and Miron, P.} [2019].  {Building a
  Maxey--Riley framework for surface ocean inertial particle dynamics}. {\em
  Phys. Fluids\/} 31, 096602.

\bibitem[Bertola {\rm et~al.}(2020)Bertola, Boehm, Putman, Xue, Robinson,
  Harris, Baldwin, Overcast and Hickerson]{Bertola-etal-20}
{\rm Bertola, L.~D., Boehm, J.~T., Putman, N.~F., Xue, A.~T., Robinson, J.~D.,
  Harris, S., Baldwin, C.~C., Overcast, I. and Hickerson, M.~J.} [2020].
  Asymmetrical gene flow in five co-distributed syngnathids explained by ocean
  currents and rafting propensity. {\em Proceedings of the Royal Society B\/}
  287, 20200657.

\bibitem[van~der Boog {\rm et~al.}(2019)van~der Boog, de~Jong, Scheidat,
  Leopold, Geelhoed, Schulz, Dijkstra, Pietrzak and
  Katsman]{vanderBoog-etal-19}
{\rm van~der Boog, C.~G., de~Jong, M.~F., Scheidat, M., Leopold, M.~F.,
  Geelhoed, S. C.~V., Schulz, K., Dijkstra, H.~A., Pietrzak, J.~D. and Katsman,
  C.~A.} [2019].  Hydrographic and biological survey of a surface-intensified
  anticyclonic eddy in the {Caribbean Sea}. {\em Journal of Geophysical
  Research\/} 124~(8), 6235--6251.

\bibitem[Breivik {\rm et~al.}(2013)Breivik, Allen, Maisondieu and
  Olagnon]{Breivik-etal-13}
{\rm Breivik, O., Allen, A.~A., Maisondieu, C. and Olagnon, M.} [2013].
  Advances in search and rescue at sea. {\em Ocean Dynamics\/} 63, 83--88.

\bibitem[Carnevale {\rm et~al.}(1991)Carnevale, Kloosterziel and {van
  H}eist]{Carnevale-etal-91b}
{\rm Carnevale, G.~F., Kloosterziel, R.~C. and {van H}eist, G. J.~F.} [1991].
  Propagation of barotropic vortices over topography in a rotating tank. {\em
  J. Fluid Mech.\/} 255, 119--139.

\bibitem[Chelton {\rm et~al.}(1998)Chelton, {deSzoeke}, Schlax, {El Naggar} and
  Siwertz]{Chelton-etal-98}
{\rm Chelton, D.~B., {deSzoeke}, R.~A., Schlax, M.~G., {El Naggar}, K. and
  Siwertz, N.} [1998].  Geographical variability of the first baroclinic rossby
  radius of deformation. {\em J. Phys. Oceanogr.\/} 28, 433--460.

\bibitem[Crisan {\rm et~al.}(2021)Crisan, Holm, Luesink, Mensah and
  Pan]{Crisan-etal-21}
{\rm Crisan, D., Holm, D.~D., Luesink, E., Mensah, P.~R. and Pan, W.} [2021].
  Theoretical and computational analysis of the thermal quasi-geostrophic
  model. arXiv:2106.14850.

\bibitem[Gouzien {\rm et~al.}(2017)Gouzien, Lahaye, Zeitlin and
  Dubos]{Gouzien-etal-17}
{\rm Gouzien, E., Lahaye, N., Zeitlin, V. and Dubos, T.} [2017].  Thermal
  instability in rotating shallow water with horizontal temperature/density
  gradients. {\em Physics of Fluids\/} 29, 101702.

\bibitem[Haller(2015)]{Haller-15}
{\rm Haller, G.} [2015].  Lagrangian coherent structures. {\em Ann. Rev. Fluid
  Mech.\/} 47, 137--162.

\bibitem[Haller and Beron-Vera(2012)]{Haller-Beron-12}
{\rm Haller, G. and Beron-Vera, F.~J.} [2012].  Geodesic theory of transport
  barriers in two-dimensional flows. {\em Physica D\/} 241, 1680--1702.

\bibitem[Haller and Beron-Vera(2013)]{Haller-Beron-13}
{\rm Haller, G. and Beron-Vera, F.~J.} [2013].  {Coherent Lagrangian vortices:
  The black holes of turbulence}. {\em J. Fluid Mech.\/} 731, R4.

\bibitem[Haller and Beron-Vera(2014)]{Haller-Beron-14}
{\rm Haller, G. and Beron-Vera, F.~J.} [2014].  {Addendum to `Coherent
  Lagrangian vortices: The black holes of turbulence'}. {\em J. Fluid Mech.\/}
  755, R3.

\bibitem[Haller {\rm et~al.}(2016)Haller, Hadjighasem, Farazmand and
  Huhn]{Haller-etal-16}
{\rm Haller, G., Hadjighasem, A., Farazmand, M. and Huhn, F.} [2016].  Defining
  coherent vortices objectively from the vorticity. {\em J. Fluid Mech.\/} 795,
  136--173.

\bibitem[Haller {\rm et~al.}(2018)Haller, Karrasch and
  Kogelbauer]{Haller-etal-18}
{\rm Haller, G., Karrasch, D. and Kogelbauer, F.} [2018].  Material barriers to
  diffusive and stochastic transport. {\em Proceedings of the National Academy
  of Sciences\/} 115, 9074--9079.

\bibitem[Holm {\rm et~al.}(2020)Holm, Luesink and Pan]{Holm-etal-21}
{\rm Holm, D.~D., Luesink, E. and Pan, W.} [2020].  Stochastic mesoscale
  circulation dynamics in the thermal ocean. {\em Phys. Fluids\/} 33, 046603.

\bibitem[Huang {\rm et~al.}(2021)Huang, Liang, Zhu, Liu and
  Weisberg]{Huang-etal-21}
{\rm Huang, M., Liang, X., Zhu, Y., Liu, Y. and Weisberg, R.~H.} [2021].
  Eddies connect the tropical {Atlantic Ocean} and the {Gulf of Mexico}. {\em
  Geophysical Research Letters\/} 48, e2020GL091277.

\bibitem[Karrasch {\rm et~al.}(2014)Karrasch, Huhn and
  Haller]{Karrasch-etal-14}
{\rm Karrasch, D., Huhn, F. and Haller, G.} [2014].  {Automated detection of
  coherent Lagrangian vortices in two-dimensional unsteady flows}. {\em Proc.
  Royal Soc. A\/} 471, 20140639.

\bibitem[Karrasch and Schilling(2020)]{Karrasch-Schilling-20}
{\rm Karrasch, D. and Schilling, N.} [2020].  Fast and robust computation of
  coherent lagrangian vortices on very large two-dimensional domains. {\em The
  SMAI journal of computational mathematics\/} 6, 101--124.

\bibitem[Kurganov {\rm et~al.}(2020)Kurganov, Liu and
  Zeitlin]{Kurganov-etal-20}
{\rm Kurganov, A., Liu, Y. and Zeitlin, V.} [2020].  Moist-convective thermal
  rotating shallow water model. {\em Physics of Fluids\/} 32~(6), 066601.

\bibitem[{Le Traon} {\rm et~al.}(1998){Le Traon}, Nadal and
  Ducet]{LeTraon-etal-98}
{\rm {Le Traon}, P.~Y., Nadal, F. and Ducet, N.} [1998].  An improved mapping
  method of multisatellite altimeter data. {\em J. Atmos. Oceanic Technol.\/}
  15, 522--534.

\bibitem[Miron {\rm et~al.}(2020{\natexlab{{\rm a}}})Miron, Medina, Olascaoaga
  and Beron-Vera]{Miron-etal-20-POF}
{\rm Miron, P., Medina, S., Olascaoaga, M.~J. and Beron-Vera, F.~J.}
  [2020{\natexlab{{\rm a}}}].  {Laboratory verification of a Maxey--Riley
  theory for inertial ocean dynamics}. {\em Phys. Fluids\/} 32, 071703.

\bibitem[Miron {\rm et~al.}(2020{\natexlab{{\rm b}}})Miron, Olascoaga,
  Beron-Vera, {Tri\~nanes}, Putman, Lumpkin and Goni]{Miron-etal-20-GRL}
{\rm Miron, P., Olascoaga, M.~J., Beron-Vera, F.~J., {Tri\~nanes}, J., Putman,
  N.~F., Lumpkin, R. and Goni, G.~J.} [2020{\natexlab{{\rm b}}}].  {Clustering
  of marine-debris-and \emph{Sargassum}-like drifters explained by inertial
  particle dynamics}. {\em Geophys. Res. Lett.\/} 47, e2020GL089874.

\bibitem[Morrison(1998)]{Morrison-98}
{\rm Morrison, P.~J.} [1998].  Hamiltonian description of the ideal fluid. {\em
  Rev. Mod. Phys.\/} 70, 467--521.

\bibitem[Morrow {\rm et~al.}(2004)Morrow, Birol and Griffin]{Morrow-etal-04a}
{\rm Morrow, R., Birol, F. and Griffin, D.} [2004].  Divergent pathways of
  cyclonic and anti-cyclonic ocean eddies. {\em Geophys. Res. Lett.\/} 31,
  L24311.

\bibitem[Nof(1981)]{Nof-81a}
{\rm Nof, D.} [1981].  On the $\beta$-induced movement of isolated baroclinic
  eddies. {\em J. Phys. Oceanogr.\/} 11, 1662--1672.

\bibitem[Olascoaga {\rm et~al.}(2020)Olascoaga, Beron-Vera, Miron,
  {Tri\~nanes}, Putman, Lumpkin and Goni]{Olascoaga-etal-20}
{\rm Olascoaga, M.~J., Beron-Vera, F.~J., Miron, P., {Tri\~nanes}, J., Putman,
  N.~F., Lumpkin, R. and Goni, G.~J.} [2020].  Observation and quantification
  of inertial effects on the drift of floating objects at the ocean surface.
  {\em Phys. Fluids\/} 32, 026601.

\bibitem[Paraguay-Delgado {\rm et~al.}(2020)Paraguay-Delgado, Carreno-Gallardo,
  Estrada-Guel, Zabala-Arceo, Martinez-Rodriguez and
  Lardizabal-Gutierre]{Paraguay-etal-20}
{\rm Paraguay-Delgado, F., Carreno-Gallardo, C., Estrada-Guel, I.,
  Zabala-Arceo, A., Martinez-Rodriguez, H.~A. and Lardizabal-Gutierre, D.}
  [2020].  {Pelagic \emph{Sargassum} spp. capture CO$_2$ and produce calcite}.
  {\em Environ Sci. Pollut. Res.\/} 42,
  https://doi.org/10.1007/s11356--020--08969--w.

\bibitem[Resiere {\rm et~al.}(2018)Resiere, Valentino, Neviere, Banydeen,
  Gueye, Florentin, Cabie, Lebrun, Megarbane, Guerrier and
  Mehdaoui]{Resiere-etal-18}
{\rm Resiere, D., Valentino, R., Neviere, R., Banydeen, R., Gueye, P.,
  Florentin, J., Cabie, A., Lebrun, T., Megarbane, B., Guerrier, G. and
  Mehdaoui, H.} [2018].  {\emph{Sargassum} seaweed on Caribbean islands: an
  international public health concern}. {\em The Lancet\/} 392, 2691.

\bibitem[Richardson(2005)]{Richardson-05}
{\rm Richardson, P.} [2005].  {Caribbean Current and eddies as observed by
  surface drifters}. {\em Deep Sea Research Part II: Topical Studies in
  Oceanography\/} 52, 429--463.

\bibitem[Ripa(1991)]{Ripa-JFM-91}
{\rm Ripa, P.} [1991].  General stability conditions for a multi-layer model.
  {\em J. Fluid Mech.\/} 222, 119--137.

\bibitem[Ripa(1993)]{Ripa-GAFD-93}
{\rm Ripa, P.} [1993].  Conservation laws for primitive equations models with
  inhomogeneous layers. {\em Geophys. Astrophys. Fluid Dyn.\/} 70, 85--111.

\bibitem[Ripa(1996)]{Ripa-RMF-96}
{\rm Ripa, P.} [1996].  Low frequency approximation of a vertically integrated
  ocean model with thermodynamics. {\em Rev. Mex. Fis.\/} 42, 117--135.

\bibitem[Simmons and Nof(2002)]{Simmons-Nof-02}
{\rm Simmons, H. and Nof, D.} [2002].  The squeezing of eddies through gaps.
  {\em J. Phys. Oceanogr.\/} 32, 314--335.

\bibitem[Smetacek and Zingone(2013)]{Smetacek-Zingone-13}
{\rm Smetacek, V. and Zingone, A.} [2013].  Green and golden seaweed tides on
  the rise. {\em Nature\/} 504, 84--88.

\bibitem[Thiffeault and Morrison(2000)]{Thiffeault-Morrison-00}
{\rm Thiffeault, J.-L. and Morrison, P.~J.} [2000].  {Classification and
  Casimir invariants of Lie--Poisson brackets}. {\em Physica D\/} 136,
  205--244.

\bibitem[{Trin\~anes} {\rm et~al.}(2021){Trin\~anes}, Putman, Goni, Hu and
  Wang]{Trinanes-etal-21}
{\rm {Trin\~anes}, J., Putman, N.~F., Goni, G.~J., Hu, C. and Wang, M.} [2021].
   {Monitoring pelagic \emph{Sargassum} inundation potential for coastal
  communities}. {\em Journal of Operational Oceanography\/}
  doi:10.1080/1755876X.2021.1902682, 1--12.

\bibitem[Wang and Hu(2016)]{Wang-Hu-16}
{\rm Wang, M. and Hu, C.} [2016].  {Mapping and quantifying \emph{Sargassum}
  distribution and coverage in the Central West Atlantic using MODIS
  observations}. {\em Remote Sens. Environ.\/} 183, 350--367.

\bibitem[Wang {\rm et~al.}(2019)Wang, Hu, Barnes, Mitchum, Lapointe and
  Montoya]{Wang-etal-19}
{\rm Wang, M., Hu, C., Barnes, B., Mitchum, G., Lapointe, B. and Montoya,
  J.~P.} [2019].  {The Great Atlantic \emph{Sargassum} Belt}. {\em Science\/}
  365, 83--87.

\end{thebibliography}

\end{document}